\documentstyle[11pt]{article}
\begin{document}
\setlength{\baselineskip}{8mm}
\vspace*{-15mm}
\begin{flushright}
{\it Physics of Fluids}, in press.\\[15mm]
\end{flushright}
\begin{center}
\begin{minipage}{135mm}
\begin{center}
{\Large \bf A non-slip boundary condition\\[1mm] for lattice Boltzmann
simulations}\\[10mm]
{\large Takaji Inamuro, Masato Yoshino, and Fumimaru Ogino}\\[3mm]
{\small Department of Chemical Engineering, Kyoto University,
 Kyoto 606-01, Japan}\\[20mm]
\end{center}
A non-slip boundary condition at a wall
for the lattice Boltzmann method is presented.
In the present method unknown distribution functions at the wall
are assumed to be an equilibrium distribution function with a counter
slip velocity which is determined so that fluid velocity at the wall
is equal to the wall velocity.
Poiseuille flow and Couette flow are calculated with the nine-velocity model
to demonstrate the accuracy of the present boundary condition.
\end{minipage}
\end{center}
\newpage
%
Recently, the lattice Boltzmann (LB) method$^{1-4}$ has been used
for many kinds of
simulations of viscous flows.  In particular, the LB method has been
successfully applied to problems of fluid flows through porous media$^{5,6}$
and multiphase fluid flows.$^{7,8}$
On the other hand, as for the implementation of a non-slip boundary
condition several approaches have been proposed.$^{9-14}$
The bounce-back
boundary condition$^{9}$ has been usually used to model stationary walls.
However, it has been found that
the bounce-back boundary condition has errors
in velocity at the wall.$^{11-14}$
Skordos$^{10}$ proposed a method for calculating  particle distributions at
a boundary node from fluid variables with the gradients of the fluid
velocity.  In his method the density is assumed to be known at the
boundary.
Noble {\it et al.}$^{11}$ developed a method for calculating the density at
the boundary and the unknown components of the particle distributions.
While their method gives accurate results with the seven-velocity model,
it is not clear whether the method can be applied to other velocity models.

In this Letter, a new approach for applying the non-slip boundary condition
at the wall is presented.
For explanation we use the LB method with the BGK collision model.$^{3, 4}$
In the method the evolution of the distribution function $f_{i}({\bf x}, t)$
of particles with velocity ${\bf c}_{i}$ at the point ${\bf x}$ and at time
$t$
is computed by the following equation:
\begin{equation}
f_{i}({\bf x}+{\bf c}_{i}\Delta t, t+\Delta t)
      - f_{i}({\bf x}, t) = - \frac{1}{\tau} [f_{i}({\bf x}, t)
                                -f_{i}^{{\rm eq}}({\bf x}, t)],
\end{equation}
where $f_{i}^{{\rm eq}}({\bf x}, t)$ is an equilibrium distribution function,
$\tau$ is a single relaxation time,
and $\Delta t$ is a time step during which the particles travel a
grid spacing.
The fluid mass density $\rho$ and the fluid velocity ${\bf u}$
are defined
in
terms of the particle distribution function by
\begin{equation}
\rho = \sum_{i} f_{i},
\end{equation}
\begin{equation}
{\bf u} = \frac{1}{\rho} \sum_{i} f_{i} {\bf c}_{i}.
\end{equation}
Here, we use the nine-velocity model$^{4, 15}$
to explain the procedure of the method,
but it is straightforward to apply the method to other velocity models.
The nine-velocity model has velocity vectors,
${\bf c}_{0} = {\bf 0}$,
${\bf c}_{i}^{I} = [ {\rm cos}(\pi (i-1)/2), {\rm sin}(\pi (i-1)/2) ]$,
and
${\bf c}_{i}^{II} = \sqrt{2}[ {\rm cos}(\pi (i-\frac{1}{2})/2),
{\rm sin}(\pi (i-\frac{1}{2})/2) ]$
for $i=1,\cdots,4$.
An equilibrium distribution function of this model
is given by
\begin{eqnarray}
f_{0}^{{\rm eq}} & = & \frac{4}{9} \rho
                (1-\frac{3}{2}{\bf u}^{2}) ,  \\
f_{i}^{I,{\rm eq}} & = & \frac{\rho }{9}
 \left[1+3({\bf c}_{i}^{I} \cdot {\bf u})
   +\frac{9}{2}({\bf c}_{i}^{I} \cdot {\bf u})^{2}
   -\frac{3}{2}{\bf u}^{2}\right] , \\
f_{i}^{II,{\rm eq}} & = & \frac{\rho }{36}
 \left[1+3({\bf c}_{i}^{II} \cdot {\bf u})
   +\frac{9}{2}({\bf c}_{i}^{II} \cdot {\bf u})^{2}
   -\frac{3}{2}{\bf u}^{2}\right] .
\end{eqnarray}

At the non-slip wall we must specify the distribution functions of
the particles whose velocity points to fluid region.
In Fig. 1 the distribution
functions $f_{2}^{I}$, $f_{1}^{II}$, and $f_{2}^{II}$ are unknowns.
In the kinetic theory of gases
the assumption of diffuse reflection is often
used at the wall.  In this approximation gas molecules that strike the wall
are assumed to leave it with a Maxwellian velocity distribution having
the velocity and the temperature of the wall.
  In general, the velocity along the wall
obtained with this assumption is not equal to that of the wall velocity,
although the normal velocity to the wall is equal to that of the wall
velocity.
The difference between the fictitious velocity and the wall velocity is
called the slip velocity.
The idea of the present method is that the unknown distribution functions
are assumed to be an equilibrium distribution function with a counter slip
velocity which is determined
so that the fluid velocity at the wall is equal to
the wall velocity.  That is, in the case of Fig. 1
the unknown distribution functions
$f_{2}^{I}$, $f_{1}^{II}$, and $f_{2}^{II}$
are assumed to be
\begin{eqnarray}
f_{2}^{I} & = & \frac{1}{9}\rho ' \left[1 + 3v_{{\rm w}}
+ \frac{9}{2}v_{{\rm w}}^{2}
             - \frac{3}{2}[(u_{{\rm w}}+u')^{2}+v_{{\rm w}}^{2}]\right], \\
f_{1}^{II} & = & \frac{1}{36}\rho ' \left[1 + 3(u_{{\rm w}}+u'+v_{{\rm w}})
             + \frac{9}{2}(u_{{\rm w}}+u'+v_{{\rm w}})^{2} \right.\nonumber \\
 & & \hspace*{15mm}\left. - \frac{3}{2}[(u_{{\rm w}}+u')^{2}
 +v_{{\rm w}}^{2}]\right], \\
f_{2}^{II} & = & \frac{1}{36}\rho ' \left[1 + 3(-u_{{\rm w}}-u'+v_{{\rm w}})
             + \frac{9}{2}(-u_{{\rm w}}-u'+v_{{\rm w}})^{2}\right. \nonumber \\
 & & \hspace*{15mm}\left.  - \frac{3}{2}[(u_{{\rm w}}+u')^{2}
 +v_{{\rm w}}^{2}]\right],
\end{eqnarray}
where $u_{{\rm w}}$ and $v_{{\rm w}}$ are the $x$ and $y$ components
of the wall velocity,
and $\rho '$ and $u'$
are unknown parameters.
The unknown $u'$ is the above-mentioned counter slip velocity.
It is noted that we have no normal velocity jump at the wall because
as mentioned above there exists no difference of the normal velocity to
the wall between the fluid and the wall on the assumption of diffuse
reflection.
The two unknown parameters are determined on
the condition that the fluid velocity at the wall is equal
to the wall velocity.
Thus, we obtain two equations corresponding to the $x$ and $y$ components of
the fluid velocity.  Moreover, the density at the wall, $\rho_{{\rm w}}$,
is an unknown quantity and is calculated by Eq. (2).  Therefore, we finally
obtain
three equations for the three unknowns.
The solutions are as follows:
\begin{eqnarray}
\rho_{{\rm w}} & = & \frac{1}{1-v_{{\rm w}}}\left[
           f_{0}+f_{1}^{I}+f_{3}^{I}+2(f_{4}^{I}+f_{3}^{II}+f_{4}^{II})
           \right], \\
\rho ' & = & 6 \frac{\rho_{{\rm w}} v_{{\rm w}}+(f_{4}^{I}
+f_{3}^{II}+f_{4}^{II})}
                {1+3v_{{\rm w}}+3v_{{\rm w}}^{2}}, \\
u' & = & \frac{1}{1+3v_{{\rm w}}}
   \left[ 6 \frac{\rho_{{\rm w}} u_{{\rm w}}-(f_{1}^{I}-f_{3}^{I}
   +f_{4}^{II}-f_{3}^{II})}
                    {\rho '}\right. \nonumber \\
 & & \hspace*{50mm}\left. -u_{{\rm w}}-3u_{{\rm w}}v_{{\rm w}}
     \frac{}{}    \right].
\end{eqnarray}

\noindent
It is noted that the same procedure can be applied to other velocity models
in spite of the number of particle velocities.
In general, for isothermal models in D-dimensional space we have D+1
constraints (D equations for each component of the velocity and one density
equation at the wall) for D unknown parameters
 ($\rho '$ and D$-$1 slip velocities) and an unknown density at
the wall.

To demonstrate the accuracy of the present boundary condition, Poiseuille flow
is calculated with the nine-velocity model.
A two-dimensional steady flow between stationary
parallel walls at $y=-L$ and $y=+L$ with a constant pressure gradient
is considered.
The pressure gradient is maintained by a density difference
between inlet and outlet.
At the inlet and the outlet the unknown distribution functions are determined
as follows.$^{16}$
At the inlet, the unknown distribution functions $f_{1}^{I}$, $f_{1}^{II}$,
and $f_{4}^{II}$ are calculated by adding a constant value to the
corresponding distribution functions at the outlet so that
$f_{1}^{I}|_{\rm in} = f_{1}^{I}|_{\rm out} + C$,
$f_{1}^{II}|_{\rm in} = f_{1}^{II}|_{\rm out} + \frac{1}{4}C$, and
$f_{4}^{II}|_{\rm in} = f_{4}^{II}|_{\rm out} + \frac{1}{4}C$
with reference to the equilibrium distribution functions given by Eqs. (5)
and (6).  The constant value $C$ is determined by setting the density
calculated by Eq. (2) to be a given value at the inlet.
The same procedure is used for calculating the unknown distribution functions
at the outlet.  In addition, the unknown distribution functions at four corners
of the inlet and the outlet are calculated as follows.  For example,
at the lower corner of inlet, $f_{1}^{I}$ and $f_{4}^{II}$ are calculated
by the above-mentioned method. Then, $f_{2}^{I}$, $f_{1}^{II}$,
and $f_{2}^{II}$ are
calculated by the present non-slip boundary condition.
 The same method is used at the other three corners.
In the following calculations
21 nodes are used between the walls.
Figure 2 shows calculated velocity profiles
$u/u_{{\rm max}}$ ($u_{{\rm max}}$ is
the velocity at $y = 0$)
with the present boundary condition
for various values of $\tau$.
It is found that the results for $0.7 \leq \tau \leq 20$
agree with the analytical solution within machine accuracy.
The values of the counter slip velocity $u'$ are
shown in Fig. 3.
It is noted that the values of $u'$ are negative for all $\tau$.
The magnitude of the counter slip velocity
increases as $\tau$ becomes larger.
Needless to mention, the magnitude of the counter slip velocity depends
on the number of nodes between the walls.
For comparison with the present results,
the results with the bounce-back boundary condition in which
$f_{2}^{I}=f_{4}^{I}, f_{1}^{II}=f_{3}^{II}$, and $f_{2}^{II}=f_{4}^{II}$
are shown in Fig. 4.  Also, the results under the condition
with $u'=0$
in the present boundary
condition, which corresponds to the usual diffuse reflection in the kinetic
theory of gases,
are shown in Fig. 5.
It is seen that both results in Fig. 4 and Fig. 5 have
slip velocities at the wall and the slip velocity increases as
$\tau$ becomes larger.  In addition, we can see that the slip velocity
under the condition with $u'=0$ in the present boundary condition
is larger than that with the bounce-back boundary condition
for the same value of $\tau$.
This is because the bounce-back boundary condition has stronger non-slip
effect than the usual diffuse reflection.
In the next problem, we calculate Couette flow to demonstrate the accuracy
of the LB method with the present boundary condition.  In the problem,
the upper plate at $y=+L$ is moved with velocity $U$ at $t>0$, and
the lower plate at $y=-L$ is at rest.  Figure 6 shows the calculated
velocity profiles $u/U$ at 200 time steps with 21 nodes between the walls
and with $\tau = 1$.  It is seen that the results agree well with
the analytical solution.  To determine the convergence rate, we perform
simulations with 11, 21, 41, and 81 nodes between the walls.
The errors are compared at the same dimensionless time corresponding to
200 time steps with 21 nodes.
Figure 7 shows the error norms $E_{1}=\sum_{y}|u-u^{*}|/\sum_{y}|u^{*}|$
and
$E_{2}=\sqrt{\sum_{y}(u-u^{*})^{2}}/\sqrt{\sum_{y}(u^{*})^{2}}$
where $u^{*}$ is the analytical solution and
the sums are taken over the same internal 9 nodes between
the walls for all cases.  The slope, $m$, of the convergence is
$m=2.0004$ for $E_{1}$, and $m=2.0006$ for $E_{2}$.
It is clearly found that the LB method with the present boundary condition
is a second-order scheme.

{}From these results, we can conclude that the present boundary condition
is accurate to model a non-slip flat boundary in the lattice Boltzmann
simulations.
It should be noted, however, that the present method has
difficulties in dealing
with corners.  One of the possible ways to handle corners is to calculate
the unknown distribution functions on each side at the corner and
to average the results of common unknown distribution functions,
while we should keep in mind that this
approach produces errors due to rounded corners.  More rigorous
considerations for corners remain in future work.
Although we use the nine-velocity model to explain the present boundary
condition, the method can be directly applied to other velocity models;
even to thermal models and to three-dimensional models.
For thermal models, we may use an equilibrium distribution function with a
counter temperature jump as well as the counter slip velocity.
The value of the
counter temperature jump can be determined by the condition for temperature
at the wall.
For three-dimensional models, we may use a counter slip velocity which
has two components on the surface of the wall.
The calculated results with the other velocity models will be reported
in the future.

\vspace{15mm}
\noindent
{\large \bf ACKNOWLEDGMENT}

This work is
supported by the Grant-in-Aid (No. 07650894) for Scientific
Research from the Ministry of Education, Science and Culture in Japan
and the General Sekiyu Research
$\&$ Development Encouragement $\&$ Assistance Foundation.

\small
\vspace*{20mm}
\noindent
\begin{description}
\item[$^{1}$] G. McNamara, and G. Zanetti, ``Use of the Boltzmann equation
to simulate lattice-gas automata,'' Phys. Rev. Lett. {\bf 61}, 2332 (1988).
\item[$^{2}$] F. Higuera and J. Jimenez, ``Boltzmann approach to lattice
gas simulations,'' Europhys. Lett. {\bf 9}, 663 (1989).
\item[$^{3}$] S. Chen, H. Chen, D. Martinez, and W.H. Matthaeus, ``Lattice
Boltzmann model for simulation of magnetohydrodynamics,'' Phys. Rev. Lett.
{\bf 67}, 3776 (1991).
\item[$^{4}$] Y.H. Qian, D. d'Humi$\grave{{\rm e}}$res, and P. Lallemand,
``Lattice BGK
models for Navier-Stokes equation,'' Europhys. Lett. {\bf 17}, 479 (1992).
\item[$^{5}$] S. Succi, E. Foti, and F. Higuera, ``Three-dimensional
flows in complex geometries with the lattice Boltzmann method,''
Europhys. Lett. {\bf 10}, 433 (1989).
\item[$^{6}$] A. Cancelliere, C. Chang, E. Foti, D.H. Rothman, and S. Succi,
``The permeability of a random medium: comparison of simulation with theory,''
Phys. Fluids A {\bf 2}, 2085 (1990).
\item[$^{7}$] A.K. Gunstensen, D.H. Rothman, and S. Zaleski, ``Lattice
Boltzmann model of immiscible fluids,'' Phys. Rev. A {\bf 43}, 4320 (1991).
\item[$^{8}$] D. Grunau, S. Chen, and K. Eggert, ``A lattice Boltzmann model
for multiphase fluids flows,'' Phys. Fluids A {\bf 5}, 2557 (1993)
\item[$^{9}$] D. d'Humi$\grave{{\rm e}}$res and P. Lallemand, ``Numerical
simulations
of hydrodynamics with lattice gas automata in two dimensions,'' Complex Syst.
 {\bf 1}, 599 (1987).
\item[$^{10}$] P.A. Skordos, ``Initial and boundary conditions for the lattice
Boltzmann method,'' Phys. Rev. E {\bf 48}, 4823 (1993).
\item[$^{11}$] D.R. Noble, S. Chen, J.G. Georgiadis, and R.O. Buckius,
``A consistent hydrodynamic boundary condition for the lattice
Boltzmann method,'' Phys. Fluids {\bf 7}, 203 (1995).
\item[$^{12}$] R. Cornubert, D. d'Humi$\grave{{\rm e}}$res, and D. Levermore,
``A Knudsen layer
theory for lattice gases,'' Physica D {\bf 47}, 241 (1991).
\item[$^{13}$] D.P. Ziegler, ``Boundary conditions for lattice Boltzmann
simulations,'' J. Stat. Phys. {\bf 71}, 1171 (1993).
\item[$^{14}$] I. Ginzbourg and P.M. Adler, ``Boundary flow condition analysis
for the three-dimensional lattice Boltzmann model,'' J. Phys. II France
{\bf 4}, 191 (1994).
\item[$^{15}$] B.T. Nadiga, ``A Study of Multi-Speed
Discrete-Velocity Gases,''
Ph. D Thesis, California Institute of Technology, 1992.
\item[$^{16}$] T. Inamuro, M. Yoshino, and F. Ogino, in preparation.
\end{description}
\newpage
\normalsize
\begin{center}
{\large \bf FIGURE CAPTION}
\end{center}

\vspace*{10mm}
\noindent
\begin{description}
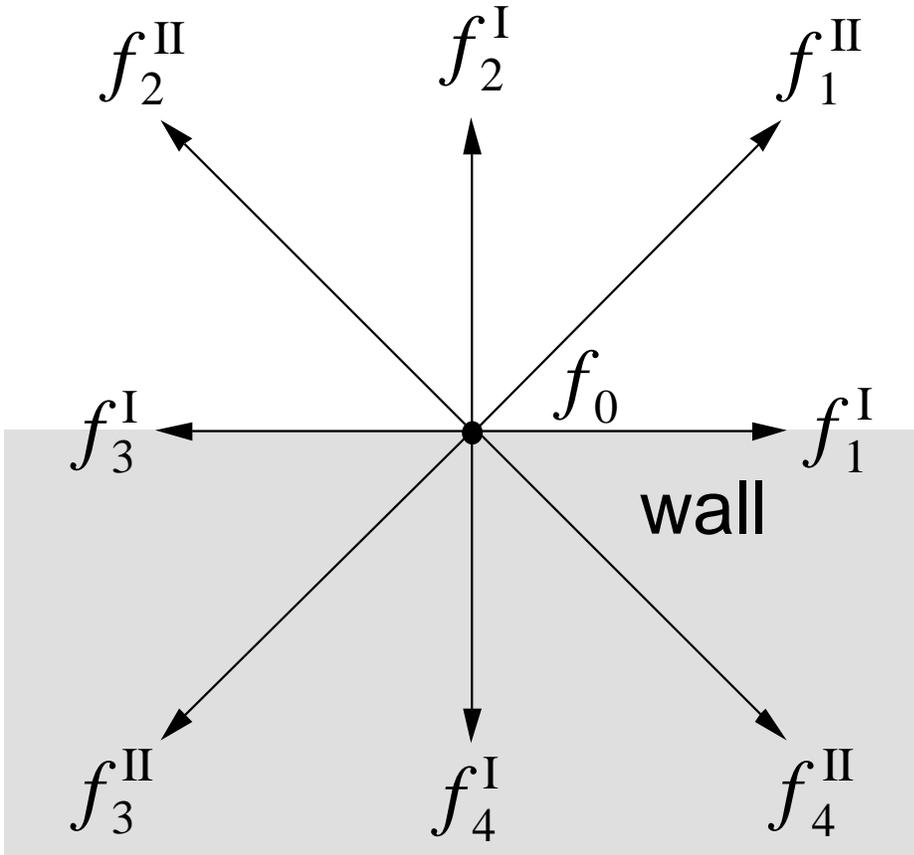
\item[{\rm Fig. 1.}] Particle distribution functions of the nine-velocity model
at the wall.\\
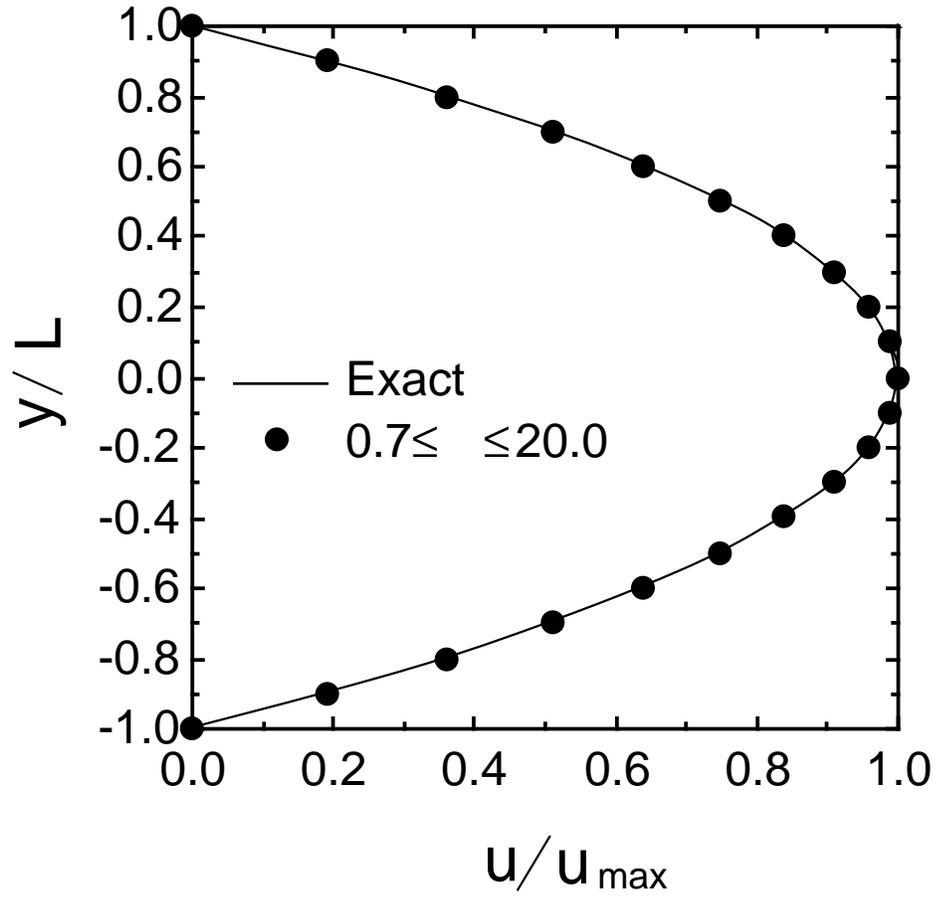
\item[{\rm Fig. 2.}] Calculated velocity profiles for Poiseuille flow
with the present boundary
condition for various values of $\tau$.\\
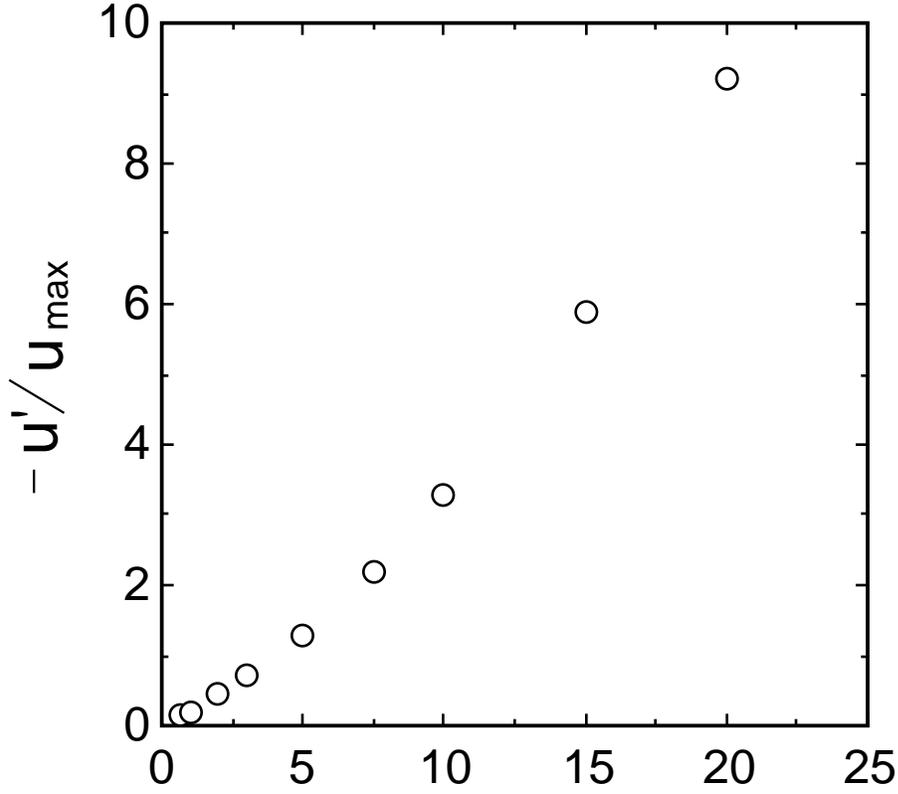
\item[{\rm Fig. 3.}] Counter slip velocity $u'$ versus single
relaxation time $\tau$ in the calculations with the present boundary
condition.\\
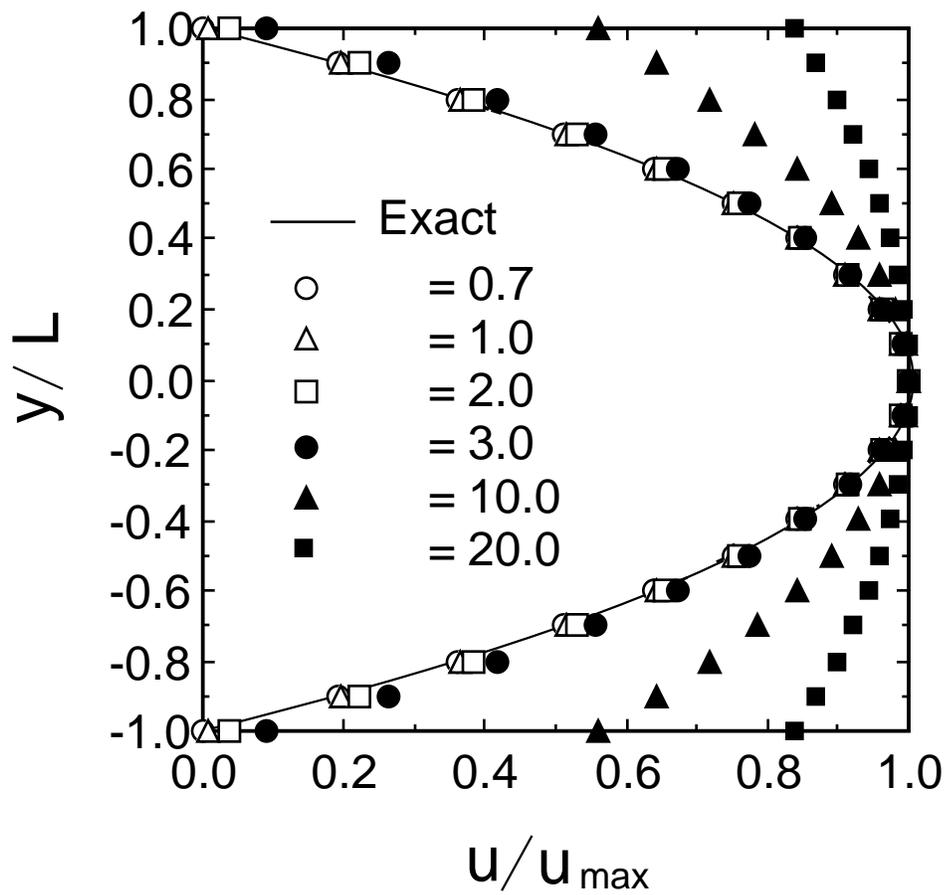
\item[{\rm Fig. 4.}] Calculated velocity profiles for Poiseuille flow
with the bounce-back
boundary condition for various values of $\tau$.\\
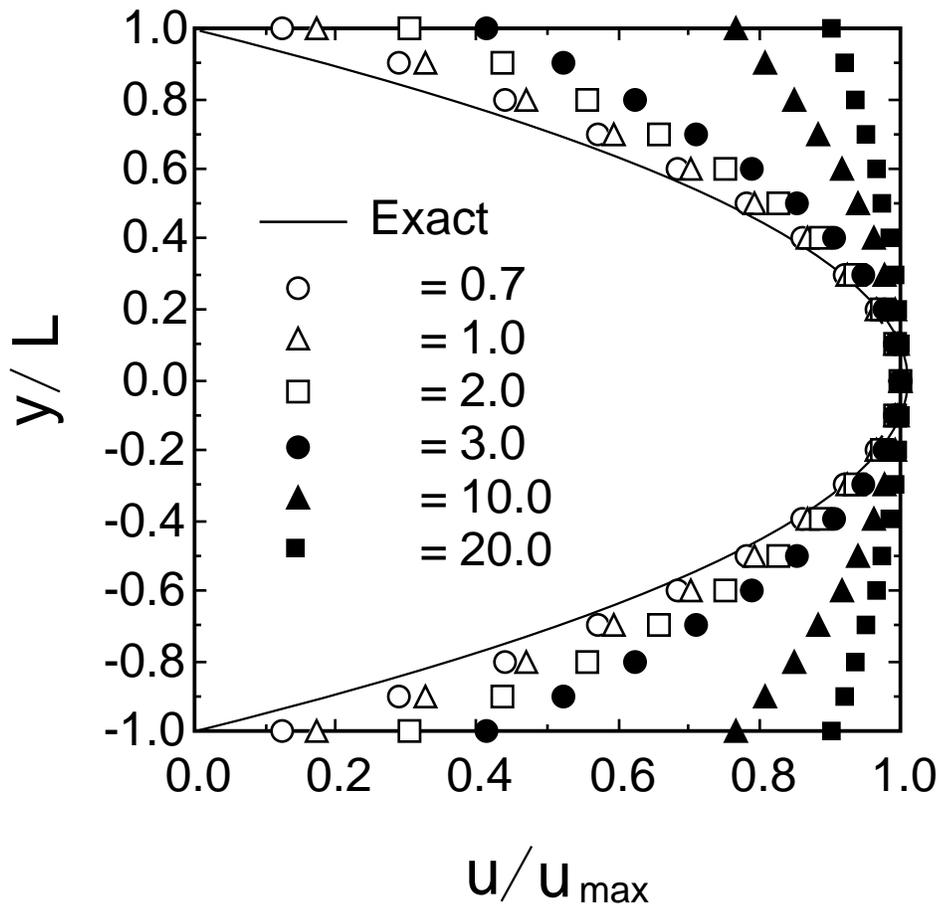
\item[{\rm Fig. 5.}] Calculated velocity profiles for Poiseuille flow
under the condition
with $u'=0$ in
the present boundary condition for various values of $\tau$.
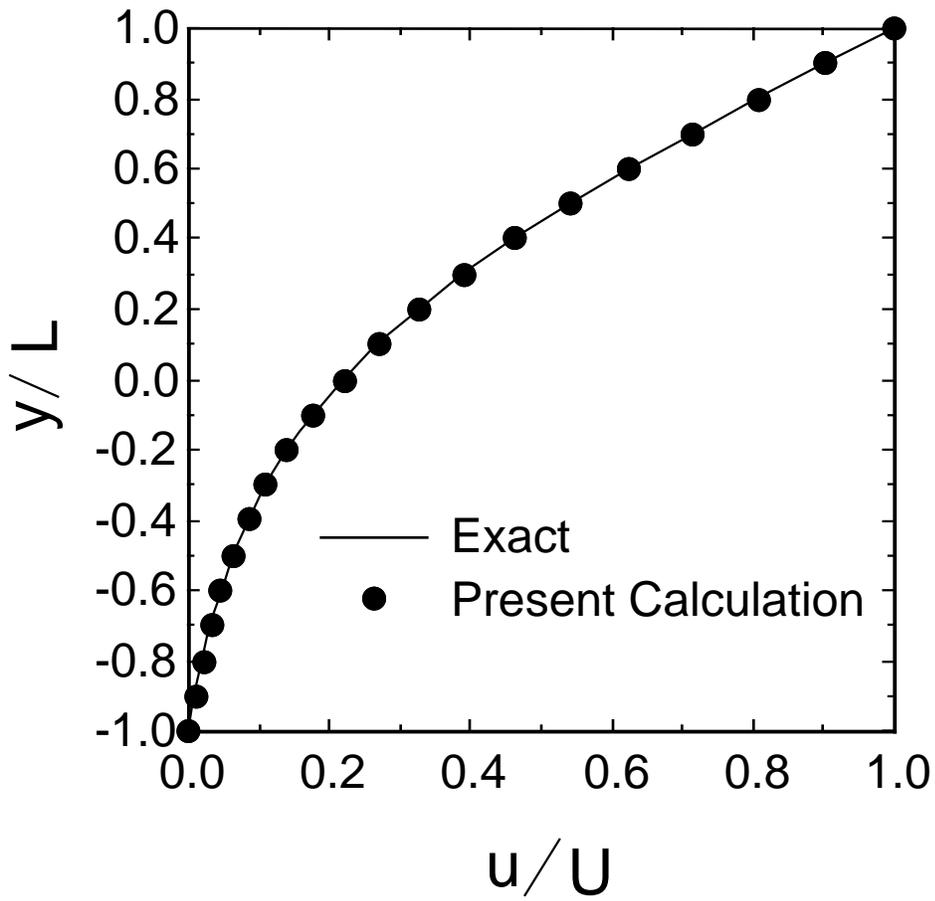
\item[{\rm Fig. 6.}] Calculated velocity profiles for Couette flow
at 200 time steps with 21 nodes between the walls and with $\tau =1$.
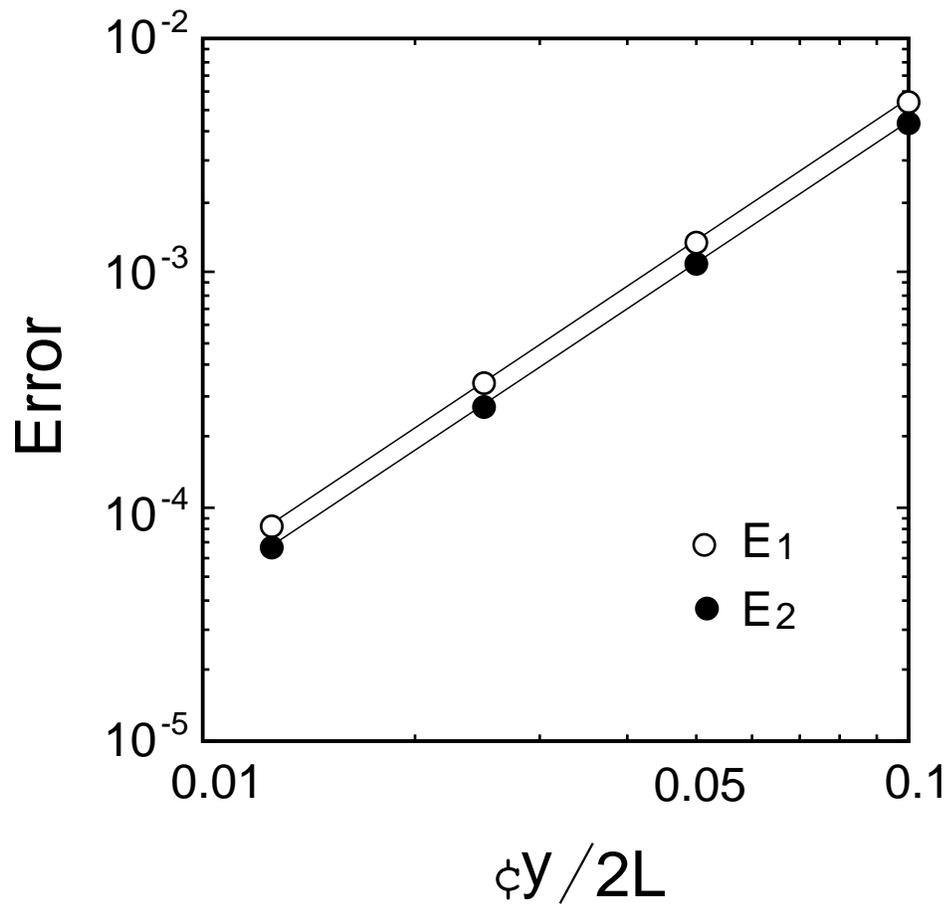
\item[{\rm Fig. 7.}] Error norms for Couette flow calculations with
11, 21, 41, and 81 nodes between the walls.
\end{description}
\end{document}